\documentclass[conference]{IEEEtran}
\IEEEoverridecommandlockouts
\usepackage{cite}
\usepackage{amsmath,amssymb,amsfonts}
\usepackage{graphicx}
\usepackage{textcomp}
\usepackage{xcolor}
\def\BibTeX{{\rm B\kern-.05em{\sc i\kern-.025em b}\kern-.08em
    T\kern-.1667em\lower.7ex\hbox{E}\kern-.125emX}}

\usepackage{algorithm}
\usepackage{algpseudocode}
\usepackage{amsthm}

\theoremstyle{remark}

\begin{document}

\title{Recursive-ARX for Grid-Edge Fault Detection
\thanks{\noindent This research is supported by the Swiss National Science Foundation under the NCCR Automation (grant agreement 51NF40\_180545)}
}
\author{
    \IEEEauthorblockN{
        Soufiane El Yaagoubi,\IEEEauthorrefmark{1} 
        Keith Moffat,\IEEEauthorrefmark{2} 
    }
    \IEEEauthorblockN{
        Eduardo Prieto Araujo,\IEEEauthorrefmark{1}
        Florian D\"orfler\IEEEauthorrefmark{2}
    }
    \\
    \IEEEauthorblockA{\IEEEauthorrefmark{1}CITCEA, 
    Polytechnic University of Catalonia, Barcelona, Spain \\ 
    Emails: soufielyaagoubi@gmail.com, eduardo.prieto-araujo@upc.edu
    }
    \IEEEauthorblockA{\IEEEauthorrefmark{2}Automatic Control Laboratory, Swiss Federal Institute of Technology (ETH), Z\"urich, Switzerland \\ 
    Emails: kmoffat@ethz.ch, dorfler@ethz.ch
    }
}

\maketitle

\begin{abstract}
Future electrical grids will require new ways to identify faults as inverters are not capable of supplying large fault currents to support existing fault detection methods and because distributed resources may feed faults from the edge of the grid. 
This paper proposes the use of real-time system identification for online power-system fault detection. 
Specifically, we implement Recursive ARX (rARX) system identification on a grid-connected inverter.
Experiments demonstrate that the proposed rARX method is able to both detect large faults quickly, and distinguish between high-impedance faults and large load increases.
These results indicate that rARX grid-edge fault detection is a promising research direction for improving the reliability and safety of modern electric grids.
\end{abstract}


\section{Introduction}

As inverters continue to replace synchronous generators,
traditional fault-detection and protection methods for electric grids may become insufficient as 
1) inverters are not capable of supplying large fault currents to support existing fault detection methods, 
2) distributed, grid-edge resources transition from grid-following to grid-forming operation \cite{b0}, and thus distributed resources may feed faults from the edge of the grid, and 
3) high-impedance faults (HIFs), which can ignite fires, can be difficult to distinguish from large load increases.

HIFs have become a particularly important challenge in arid environments in which faults can lead to large wildfires.
HIFs occur when a distribution network conductor comes into contact with an object such as a tree or a pole that provides a high-impedance path to ground. As the resulting fault current can be lower than the nominal max current at the fault location \cite{b1}, conventional protection systems have a difficult time detecting the HIF and tripping the relay \cite{b2}. Studies indicate that approximately/at least 20\% of the faults in distribution systems are HIFs \cite{b3,b4}. 

\begin{figure}
\centering
    \includegraphics[scale=0.25]{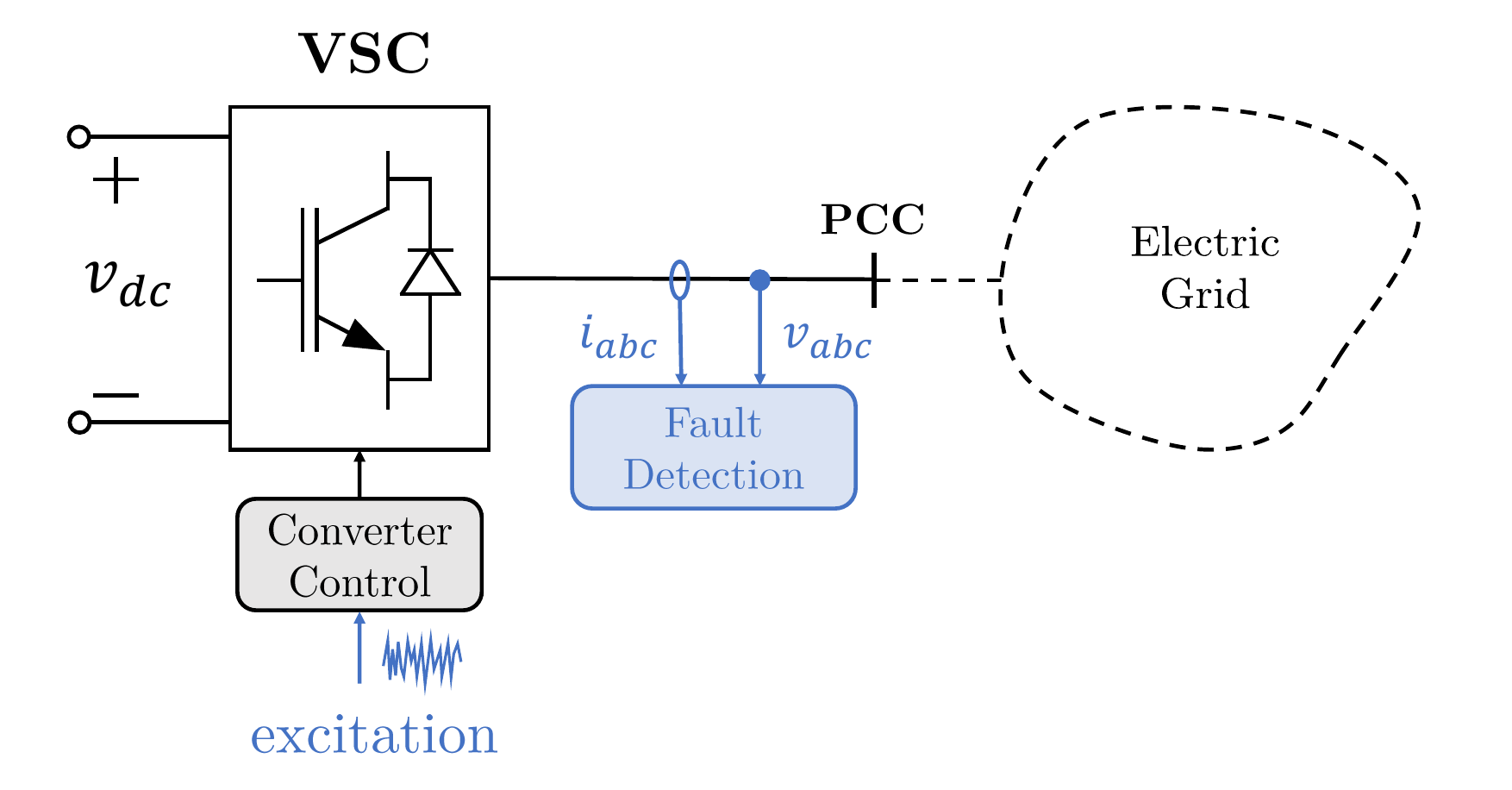}
    \caption{Data-Driven Grid-edge Fault Detection}
    \label{Fig1}
\end{figure}

\begin{figure*}
\centering
    \includegraphics[scale=0.53]{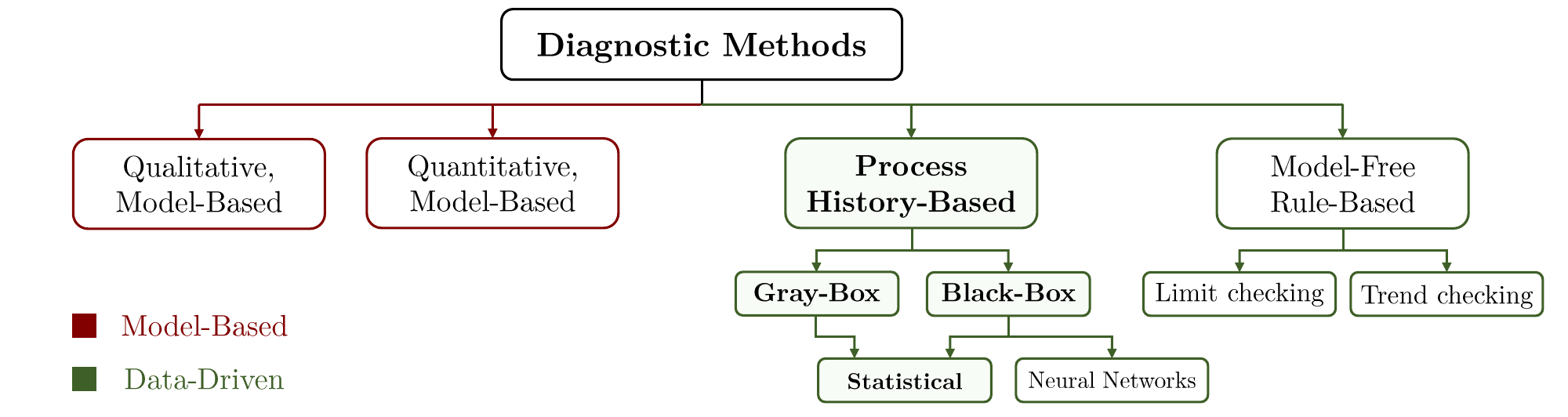}
    \caption{Classification scheme for FDD methods.}
    \label{Fig2}
\end{figure*}

Fig. \ref{Fig2} provides a classification scheme for Fault Detection and Diagnosis (FDD), dividing the methods into model-based or data-driven methods categories. Model-based approaches are either quantitative and qualitative models \cite{b15}. While these methods can be powerful, they require extensive system-specific modeling and knowledge, making them impractical in many real-world applications \cite{b11}. 

In contrast, data-driven methods are divided into process history-based and model-free rule-based approaches \cite{b15}. Model-free rule-based methods, such as limit-checking and trend analysis, provide a simple fault detection strategy based on predefined thresholds such as voltage lower-limits. While these methods are sufficient for detecting low-impedance faults (LIFs) \cite{b01}, they can fail to detect HIFs.

Process history-based methods are further categorized into black-box and grey-box models \cite{b15}. Grey-box models incorporate some physics-based knowledge, while black-box models do not. 
Black-box models require only input/output data to identify faults by deriving patterns from the data, and thus provide a flexible, plug-and-play fault identification solution.
The literature contains both neural network-based approaches \cite{b30,b31} and statistics-based system identification approaches. Statistical approaches, which consider the noise characteristics of an underlying process and are optimal in some way, have  been used for fault detection in a number of applications such as HVAC \cite{b17} and Lithium-ion battery systems \cite{b18, din2016scalable}.

This paper proposes rARX grid-edge fault detection based on the insight that faults change grid dynamics. 
That is, if a fault occurs in the system, the dynamic relation between the current injections at a point on the grid edge and the voltage measured at that point will change. 
The rARX algorithm identifies time-domain parameters, similar to \cite{b6}, that are monitored to detect if the system dynamics have changed.
Our hypothesis, tested in this paper, is that rARX system identification can both detect LIFs quickly using no model information (black-box), and distinguish between HIFs and load increases based on specific parameters (grey-box). 

This paper is organized as follows: Section II provides the necessary background. Section III presents the fault detection approach. Section IV validates the method through simulations and comparisons and Section V concludes the paper.

\section{Background}

\subsection{ARX System Identification}
We consider a time-domain ARX model of a system with combined input and output dimension $l$.
The set of measurements for $\rho$ steps into the past at time \( k \) are collected into the data regressor \( \varphi(k) \in \mathbb{R}^{l\rho} \),
\begin{align*}
    \varphi(k) = \begin{bmatrix} 
    y(k\!-\!\!1)^\top & \!\!\!\!  \cdots & \!\!\!\!  y(k\!-\!\!\rho)^\top & \!\!\!\!  u(k\!-\!\!1)^\top & \!\!\!\!  \cdots \!\!\!\!  & u(k\!-\!\!\rho)^\top 
    \end{bmatrix}^\top \!\! \!\! .
\end{align*}

The dynamics model for a one-dimensional output is
\begin{equation}\label{arxedo}
   y(k) = \theta \varphi(k) + \epsilon(k), 
\end{equation}
where $\theta \in \mathbb{R}^{1 \times l\rho}$ is the predictor matrix
and $\epsilon(k)$ is the white noise innovation \cite{ljung1997book}.

The batch least-squares estimate of \(\theta\) at timestep $k$ minimizes the cost function
\begin{equation*}
    \mathcal{J}_N(k) = \sum_{i=1}^{N} \| y(k-i) - \theta \varphi(k-i) \|^2
\end{equation*}
for some data window-length $N$.
Instead, we implement the Recursive ARX (rARX) algorithm, which estimates \(\theta (k) \), the predictor at timestep $k$, based on the exponential forgetting factor \(\lambda \in (0,1]\): 
\begin{equation}\label{eqn:rARXcost}
    \mathcal{J}_\infty(k) = \sum_{i=0}^{\infty} \lambda^i \| y(k-i) - \theta(k) \varphi(k-i) \|^2.
\end{equation}
$\mathcal{J}_\infty(k)$ prioritizes recent measurements more than $\mathcal{J}_N(k)$, and thus is better suited for fault detection applications for which speed-of-detection is important.
The forgetting factor can be chosen explicitly, or updated automatically based on data \cite{b18}.
Given the Kalman gain $K \in \mathbb{R}^{l\rho}$ at time $k$,
\begin{equation*}
    K(k) = \frac{P(k-1)\varphi(k)}{\lambda + \varphi(k)^\top P(k-1)\varphi(k)},
\end{equation*}
calculated using the recursively-calculated covariance of $\theta(k)$,
\begin{equation*}
    P(k) = \frac{1}{\lambda} \big( P(k-1) - K(k)\varphi(k)^\top P(k-1) \big),
\end{equation*}
the recursively-calculated
\begin{equation}\label{eqn:rARXsoln}
    \theta(k) = \theta(k-1) + \big( y(k) -  \theta(k-1)\varphi(k) \big) K(k)^\top
\end{equation}
minimizes $\mathcal{J}_\infty(k)$ \cite{ljung1997book}.

\subsection{Converter Data Preprocessing}

For grid-edge system identification, as shown in Fig. \ref{Fig1}, we define the input as the current injected into the grid at the PCC, and the output as the voltage measured at the PCC.
The three-phase \( v_{abc}(k) \) and \( i_{abc}(k) \) signals are converted to \( dq \) reference-frame signals using the frequency measured locally by the converer's Phase-Locked-Loop (PLL).
The inputs and outputs used for the small-signal system idenfitication are the time-difference signals
\begin{align*}
    u(k) &=  \Delta i_{dq}(k) = i_{dq}(k) - i_{dq}(k-1) \in \mathbb{R}^2, \text{ and}\\
    y(k) &= \Delta v_{dq}(k) = v_{dq}(k) - v_{dq}(k-1) \in \mathbb{R}^2.
\end{align*}
To ensure persistency of excitation, we add random binary sequence (RBS) noise to the output of the converter's current-control loop.

\section{Recursive ARX for Fault Detection}

This section describes the grid-edge fault detection method that we propose in this paper.
The rARX algorithm estimates the grid parameters at each time step \(k\) and a fault is detected if the parameters vary significantly.  
The predictor for $y_d(k) = \Delta v_{d}(k)$ is $\boldsymbol{\theta}_d(k)$, and the predictor for $y_q(k) = \Delta v_{q}(k)$ is $\boldsymbol{\theta}_q(k)$:
\begin{align*}
    \Delta v_{d}(k) &= \boldsymbol{\theta}_d(k) \varphi(k) + \epsilon_d(k) \text{, and} \\
    \Delta v_{q}(k) &= \boldsymbol{\theta}_q(k) \varphi(k) + \epsilon_q(k).
\end{align*}
$\boldsymbol{\theta}_d(k)$ and $\boldsymbol{\theta}_q(k)$ are vertically-concatenated to give
\begin{align*}
    \boldsymbol{\theta}(k) = \begin{bmatrix}
        \boldsymbol{\theta}_d(k) \\
        \boldsymbol{\theta}_q(k)
    \end{bmatrix} \in \mathbb{R}^{2 \times 4\rho}.
\end{align*}

\begin{figure}
\centering
    \includegraphics[scale=0.46]{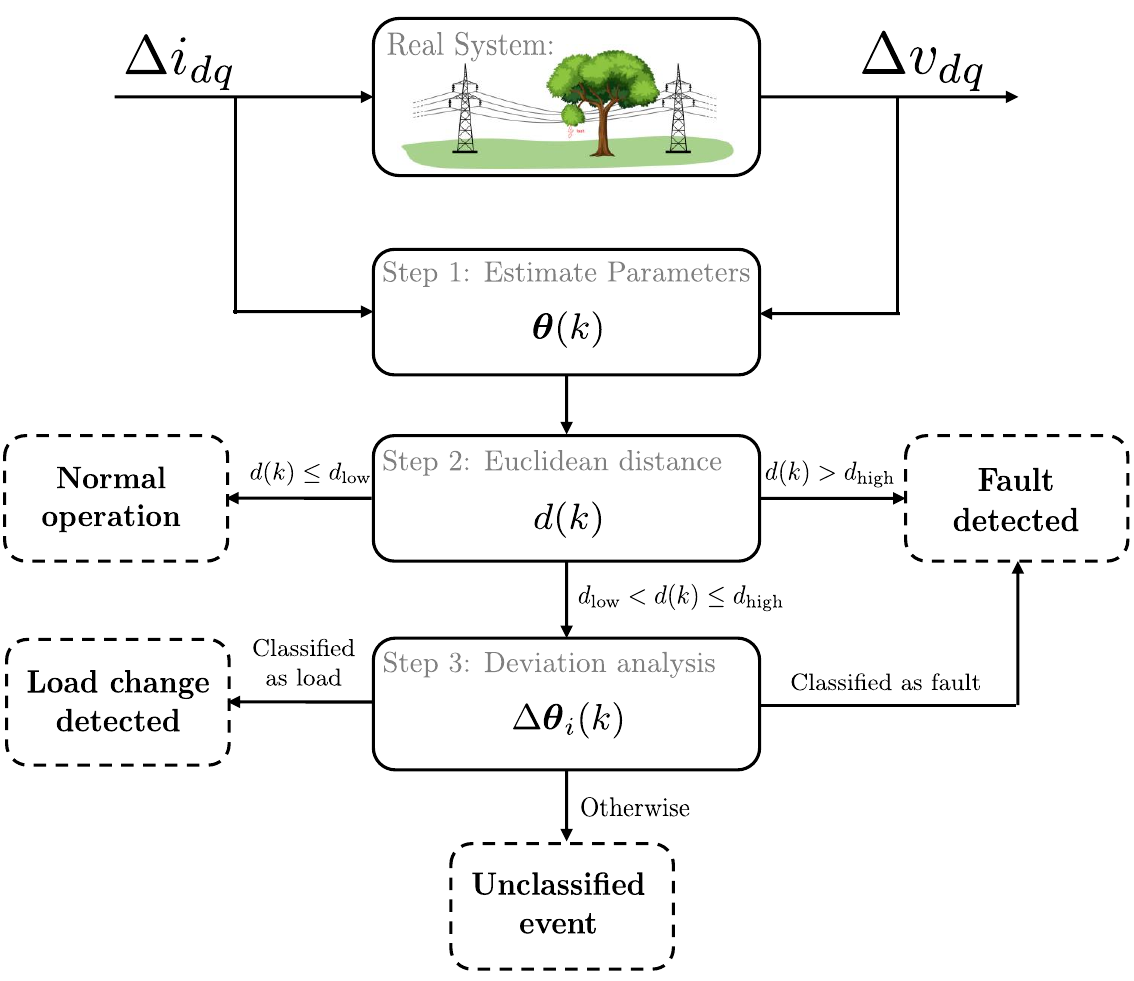}
    \caption{Proposed method for detecting faults.}
    \label{Fig4}
\end{figure}

The algorithm we propose, described in Fig. \ref{Fig4} and Algorithm \ref{alg:fault_detection}, consists of two criteria. 
Both criteria use the deviation of \(\boldsymbol{\theta}(k)\) from \(\boldsymbol{\theta}^*\), the nominal output predictor that was obtained at some point in the past at which the system was operating with no faults present.

The first, Euclidian-distance criterion, based on 
\begin{equation*}
    d(k) = \|\boldsymbol{\theta}(k) - \boldsymbol{\theta}^*\|_F, 
\end{equation*}
($\| \cdot \|_F$ denotes the Frobenius Norm) checks for large (high-admittance) faults, which change the output predictor parameters significantly. The Euclidian-distance criterion uses $d_\text{high}$, which is determined by running experiments and/or simulations offline.

The second criterion is triggered if \( d_\text{low} < d(k) \leq d_\text{high} \), and is designed to catch HIFs.
The intuition for the second criterion is that the introduction of a resistive path to ground (used to model a HIF), and an inductive path to ground (used to model an inductive load increase such as a motor-start), will produce different dynamics. This intuition is demonstrated with specific circuit parameters in Appendix \ref{app:modelAnalysis}. The parameters identified by the rARX model indicate which set of dynamics is present in the system. Standard low-voltage limit-checking methods are not able to make the same distinction.

To differentiate between HIFs and large load increases, each 
\begin{align*}
    \Delta \boldsymbol{\theta}_{d,i}(k) &= \boldsymbol{\theta}_{d,i}(k) - \boldsymbol{\theta}^*_{d,i} \text{ and} \\
    \Delta \boldsymbol{\theta}_{q,i}(k) &= \boldsymbol{\theta}_{q,i}(k) - \boldsymbol{\theta}^*_{q,i}
\end{align*}
is analyzed individually ($i$ indicates the index of the component in the $\Delta \boldsymbol{\theta}_d$ or $\Delta \boldsymbol{\theta}_q$ vector).
The deviation of each vector component is compared with a library of known faults and large load-increase behaviors, providing the basis for the fault/no fault decision.
The library is built by running experiments and/or simulations offline.
As this portion of the method requires system information (i.e., a simulator), it is grey-box rather than black-box.

\begin{algorithm}[t]
\caption{Fault Detection Algorithm}
\label{alg:fault_detection}
\begin{algorithmic}
    \State \textbf{Initialization:} 
    \State Set nominal parameters \( \boldsymbol{\theta}^* \) using fault-free system data  \\
    Set thresholds \( d_\text{high}, d_\text{low} \) and the forgetting factor \(\lambda\)
    \For{each time step \( k \)}
        \State Estimate parameters \( \boldsymbol{\theta}(k) \)
        \State Compute Euclidean distance: \( d(k) \)
        \If{ \( d(k) > d_\text{high} \) }  
            \State \textbf{Fault detected}
        \ElsIf{ \( d(k) > d_\text{low} \) }

            \State Compare $\Delta \boldsymbol{\theta}_{d,i}$ and $\Delta \boldsymbol{\theta}_{q,i}$ with recorded cases
            \If{ Deviation matches a fault case }
                \State \textbf{Fault detected}
            \ElsIf{ Deviation matches a load increase case }
                \State \textbf{Load increase detected}
            \Else
                \State \textbf{Unclassified event}
            \EndIf
        \Else
            \State \textbf{Normal operation}
        \EndIf
    \EndFor
\end{algorithmic}
\end{algorithm}

\section{Simulation Validation \& Comparison}

To validate our proposed rARX for fault detection and compare it with current methods, we use Simscape Electrical in MATLAB/Simulink to perform electromagnetic transient (EMT) simulations. The electrical parameters are provided in Table \ref{param_red}. 

\begin{figure*}
\centering
    \includegraphics[scale=0.24]{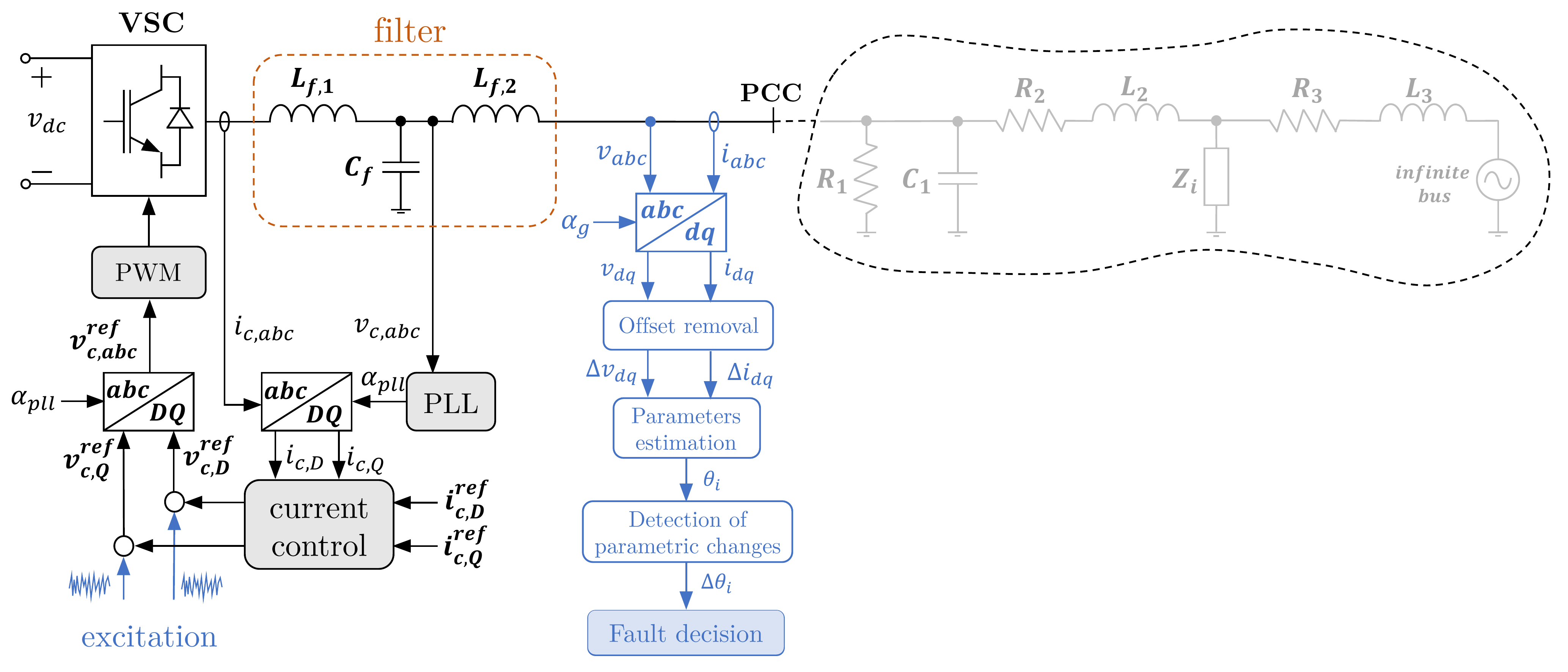}
    \caption{Fault-detection simulation set-up}
    \label{fig:simulationSetup}
\end{figure*}

Fig. \ref{fig:simulationSetup} describes the circuit used for the tests.
This circuit includes an electrical network line with load \( R_1, C_1 \) and a Thevenin-equivalent model of the grid, which includes an infinite bus and the impedances \( R_2, L_2, R_3, \text{ and } L_3 \). 
The impedance \( Z_i \) is activated at \( t_{\text{start}}\) and deactivated at \( t_{\text{end}} \) and is either a fault or a load, depending the test being run. In the case of a fault, \( Z_i = R_{\text{fault}} \) and, in the case of a load increase, \( Z_i = jX_{\text{load}} \), where \( X_{\text{load}} = \omega_g L_{\text{load}} \). 
We add two uncorrelated RBS signals to the output of the current controller to provide system excitation, each with an amplitude of \(0.1 \, \text{p.u.}\) and a sampling frequency of \(5 \, \text{kHz}\) into the control loop of the converter, as shown in Fig. \ref{fig:simulationSetup}.
$\rho = 3$ was chosen for the ARX model as it results in an $R^2$ coefficient of determination \cite{R_squared} below $5\%$, providing a good fit to the model while not over-parameterizing for the training data.

\begin{table}[H]
    \centering
    \caption{Electrical Parameters of the Numerical Experiment.}
    \scriptsize 
    \begin{tabular}{l||l|l}
    \hline
    \textbf{Parameter} & \textbf{Symbol} & \textbf{Value} \\ \hline
    Voltage, power, \& freq. base & \(V_b, S_b, f_b\) & 380 V, 1.5 kVA, 50 Hz \\ \hline
    LCL filter components & \(L_{f1}, L_{f2}, C_f\) & 0.08 p.u., 0.05 p.u., 0.08 p.u. \\ \hline
    Load component & \(R_1, C_1\) & 2 p.u., 0.05 p.u. \\ \hline
    Line 1 & \(R_2, L_2\) & 0.015 p.u., 0.15 p.u. \\ \hline
    Line 2 & \(R_3, L_3\) & 0.015 p.u., 0.15 p.u. \\ \hline
    \end{tabular}
    \label{param_red}
\end{table}

\subsection{Demonstration of the Proposed rARX Method}

\subsubsection{rARX for LIF detection}

\begin{figure}
\centering
    \includegraphics[scale=0.26]{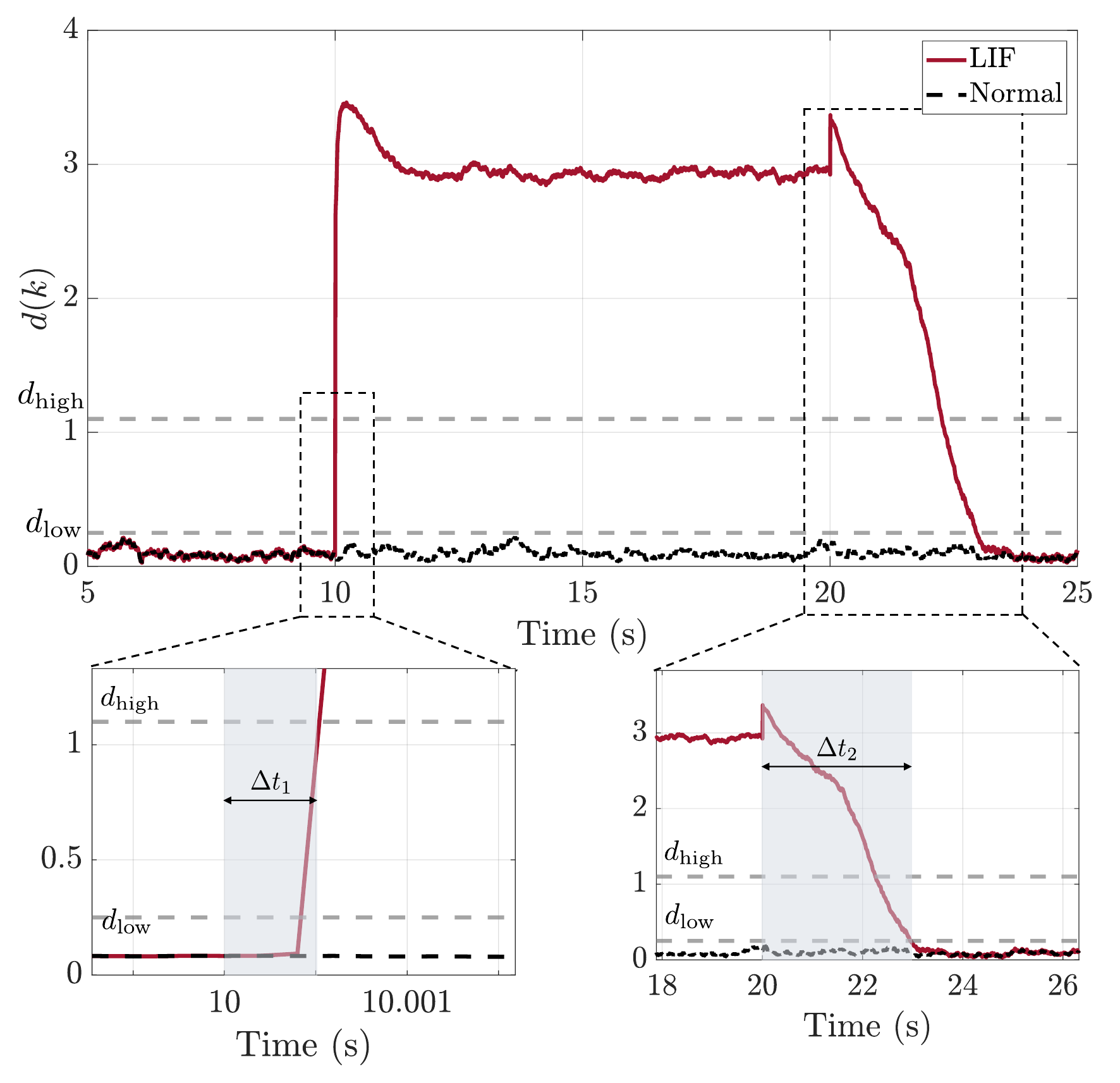}
    \caption{Low-impedance fault (LIF) test}
    \label{Fig:LIF}
\end{figure}

To demonstrate the operation of the proposed method for a LIF, a simulation is carried out in which a LIF ($R=20\Omega$) is active from \( t_{\text{start}} = 10 \, s \) to \( t_{\text{end}} = 20 \, s \). 
Fig. \ref{Fig:LIF} illustrates how the parameter vector distance $d(k)$ varies with time for the LIF. \( \Delta t_1 = 0.0005 s\) is the time elapsed from \( t_{\text{start}} \) until \( d(k) > d_\text{high} \), and \( \Delta t_2 = 2.9 s\) is the time from \( t_{\text{end}} \) until \( d(k) \leq d_\text{low} \). While noting that these detection times are dependent upon the forgetting factor parameter $\lambda$ and therefore can be easily increased/decreased, the \( \Delta t_1 = 0.0005 s\) fault detection time for the given parameters is promising.

\subsubsection{rARX for HIF detection}

To demonstrate the operation of the proposed method for a HIF, two simulations are carried out, one with a HIF, and one with a load increase. The HIF and the load increase are active from \( t_{\text{start}} = 10 \, s \) to \( t_{\text{end}} = 20 \, s \).

\begin{figure}
\centering
    \includegraphics[scale=0.26]{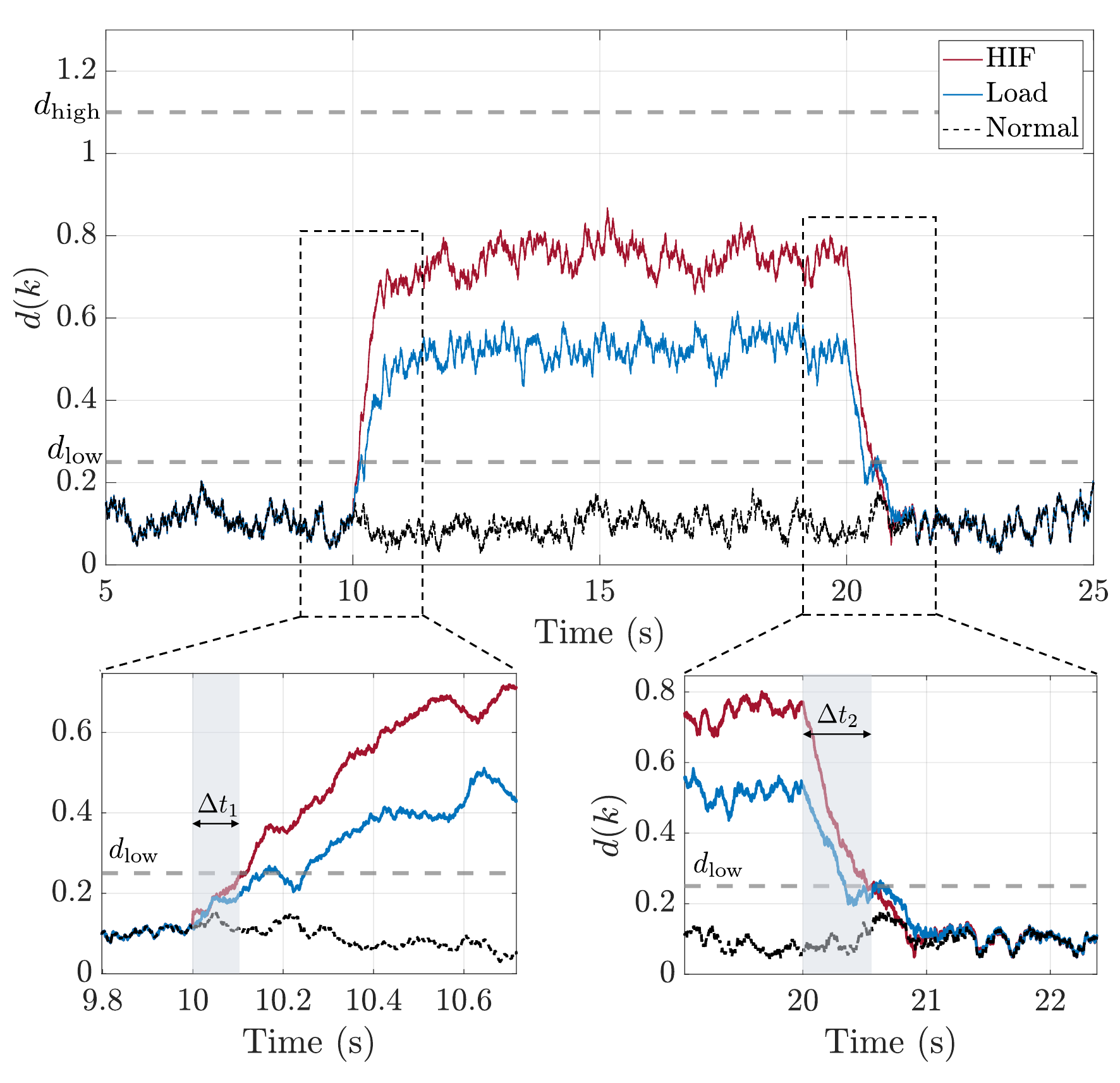}
    \caption{High-impedance fault (HIF) test}
    \label{Fig:HIF}
\end{figure}

\begin{figure}
\centering
    \includegraphics[scale=1.25]{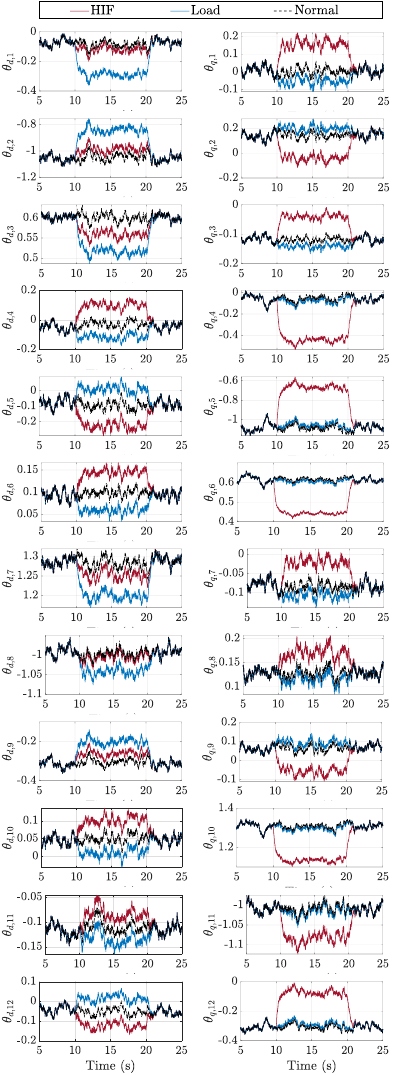}
    \caption{Parameters for the high-impedance fault test}
    \label{Fig6}
\end{figure}

Fig. \ref{Fig:HIF} illustrates how the parameter vector distance $d(k)$ varies with time for the HIF. \( \Delta t_1 = 0.10s\) is the time elapsed from \( t_{\text{start}} \) until \( d(k) > d_\text{low} \), and \( \Delta t_2 = 0.6s\) is the time from \( t_{\text{end}} \) until \( d(k) \leq d_\text{low} \).
Once again, these detection times are dependent upon the forgetting factor parameter $\lambda$, which may be adjusted as necessary.
The parameter-specific/recorded-cases criterion in Algorithm \ref{alg:fault_detection} is activated when \( d_\text{low} < d(k) \leq d_\text{high} \), which is illustrated in Fig. \ref{Fig6}. The significant differences between parameter estimates when a (resistive) HIF has occurred and when an (inductive) load increase has occurred demonstrate the viability of rARX for detecting HIFs.

\subsection{Comparison to Voltage Limit-Checking}

Voltage limit-checking is the de-facto standard for grid-edge fault detection.
The limit-checking method can be applied in the \( dq \)-reference frame using the following user-defined limits:
\begin{equation}
\begin{aligned}
v_{d,\text{min}} < v_{d}(k) < v_{d,\text{max}} \\
v_{q,\text{min}} < v_{q}(k) < v_{q,\text{max}}
\end{aligned}
\label{eqvdvq}
\end{equation}

\begin{figure}
\centering
    \includegraphics[scale=0.26]{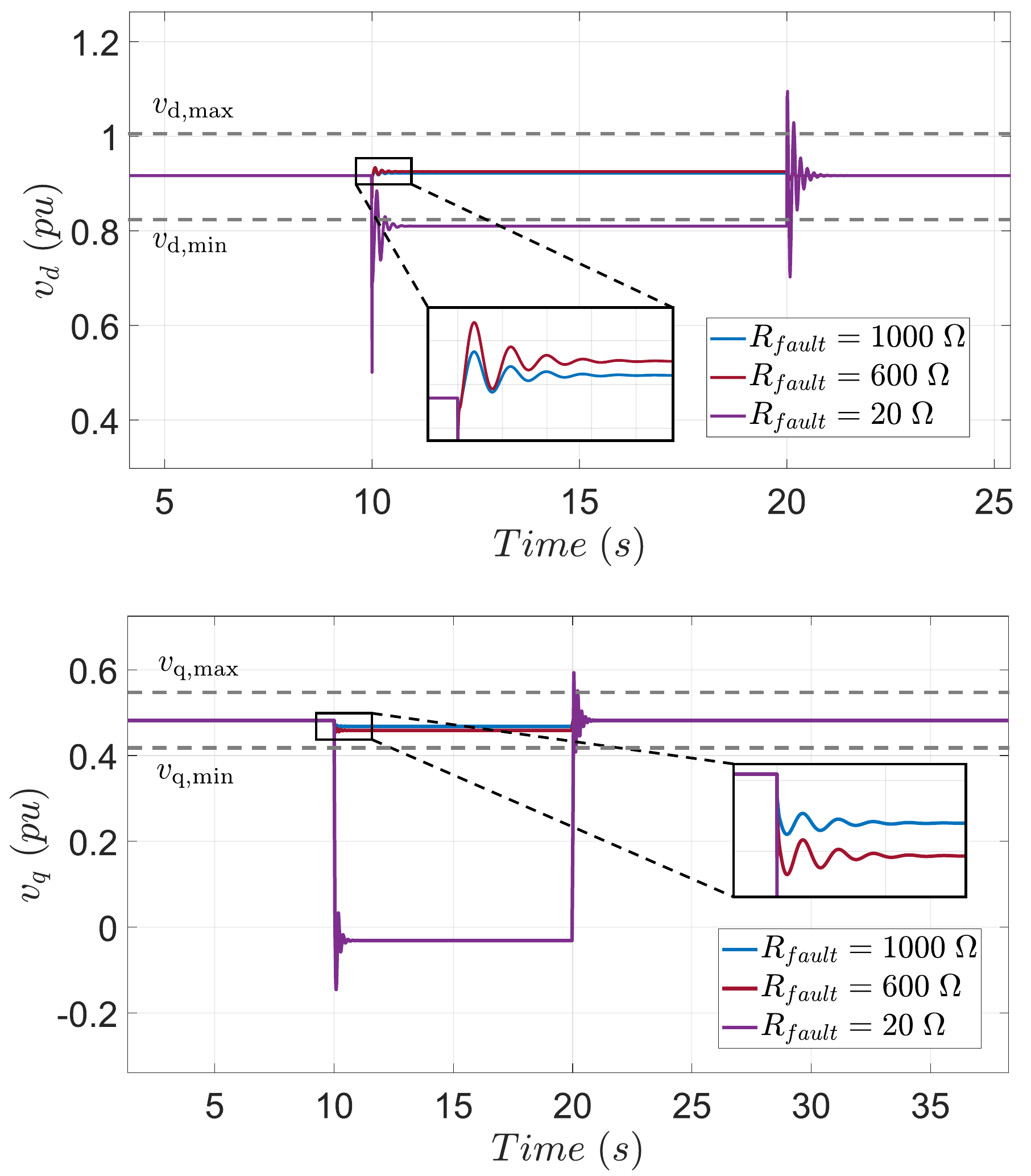}
    \caption{Demonstration of limit-checking for faults with three different resistance values}
    \label{vdq_lim}
\end{figure}

Fig. \ref{vdq_lim} shows how \(v_d\) and \(v_q\) vary for different values of \(R_{\text{fault}}\). It is evident that when the fault resistance is low, i.e. \(R_{\text{fault}} = 20 \, \Omega\), the limit-checking method can easily detect the fault. 
However when the fault resistance is high, i.e. \(R_{\text{fault}} = 600 \, \Omega\) or \(R_{\text{fault}} = 1000 \, \Omega\), the limit-checking method cannot detect the fault. 
Thus, the parameter-specific test in Step 3 of our method (Fig. \ref{Fig4}) is well motivated.

\section{Conclusion}

This paper presented a novel approach for real-time fault detection in power grids using rARX. 
The proposed method continuously updates system parameters and detects faults by evaluating deviations from the nominal model. The algorithm follows a two-step decision-making process: first, it detects anomalies based on parameter deviations, and second, it distinguishes between grid faults and inductive load increases by analyzing the specific parameters.

The results in this paper indicate that rARX grid-edge fault detection is a promising research direction. 
Future research directions include developing a method for building the fault-vs.-load-increase parameter dictionary, which is necessary for reliably detecting high-impedance faults, and increasing the speed of the fault detection by using methods such as optimal forgetting factors and excitation design.

\appendices

\section{Model-based Intuition for Differentiating Between HIFs and Inductive Load Increases}\label{app:modelAnalysis}

\begin{figure}
\centering
    \includegraphics[scale=0.23]{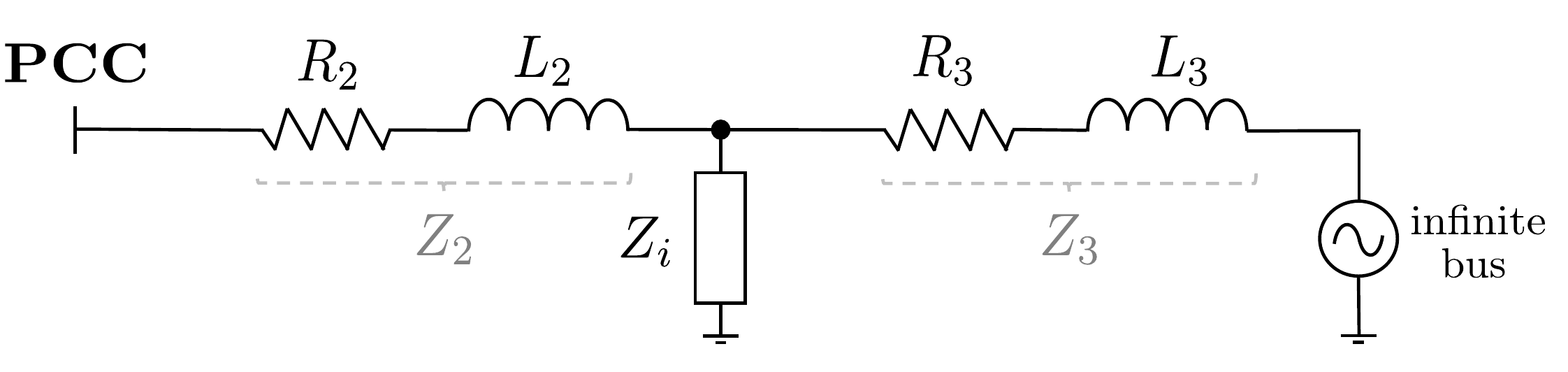}
    \caption{Equivalent circuit with a fault and a load increase.}
    \label{Zi} 
\end{figure}

This Appendix describes how the poles of the small-signal grid model depend on whether the introduced disturbance is an HIF or an inductive load.
The purpose is to demonstrate that, given sufficient data, an rARX model should be able to distinguish between an HIF and an inductive load.
The grid model is taken from the converter's perspective, at the point of common coupling, as shown in Fig. \ref{fig:simulationSetup}.
For simplicity, we remove \( R_1 \) and \( C_1 \), producing the circuit in Fig. \ref{Zi}.

The small-signal \( dq \) grid impedance, \( Z_g(s) \), relates the terminal voltages to the currents at the PCC at the steady-state grid frequency \( \omega_g \). \( Z_g(s) \) consists of four transfer functions: \( Z_{g,dd}(s) \), \( Z_{g,qd}(s) \), \( Z_{g,dq}(s) \), and \( Z_{g,qq}(s) \), such that \cite{harnefors2007modeling}
\begin{equation*}
\begin{bmatrix} \Delta v_d(s) \\ \Delta v_q(s) \end{bmatrix} =
\begin{bmatrix} Z_{g,dd}(s) & Z_{g,qd}(s) \\ Z_{g,dq}(s) & Z_{g,qq}(s) \end{bmatrix}
\begin{bmatrix} \Delta i_d(s) \\ \Delta i_q(s) \end{bmatrix}.
\end{equation*}
The grid impedance before the fault is 
\begin{equation*}
Z_{g, \text{nom}}(s) = Z_2(s) + Z_3(s),
\end{equation*}
where
\begin{equation*}
Z_2(s) = \begin{bmatrix} R_2 + s L_2 & -\omega_g L_2 \\ \omega_g L_2 & R_2 + s L_2 \end{bmatrix}
\end{equation*}
and
\begin{equation*}
Z_3(s) = \begin{bmatrix} R_3 + s L_3 & -\omega_g L_3 \\ \omega_g L_3 & R_3 + s L_3 \end{bmatrix}.
\end{equation*}
After the disturbance, the grid impedance is
\begin{equation*}\label{Zfault}
Z_{g, \text{post}}(s) = \left( Z_3^{-1}(s) + Z_{\text{i}}^{-1} \right)^{-1} + Z_2(s). 
\end{equation*}
where $Z_{\text{i}}$ is either $Z_{\text{fault}}$ or $Z_{\text{load}}$.

As HIFs typically occur when a conductor comes into contact with a short-but-high-resistance path to ground, we approximate 
\begin{equation*}
Z_{\text{fault}}(s) = \begin{bmatrix} R_{\text{fault}} & 0 \\ 0 & R_{\text{fault}} \end{bmatrix}.
\end{equation*}
The poles for each of the four transfer functions after the fault, obtained with the parameters in Table \ref{param_red}, are

\begin{equation*}\label{faultpol2}
s_{1,2,\text{fault}} = -\frac{20}{3}R_{\text{fault}} - \frac{1}{10} \pm j.
\end{equation*}

We focus on differentiating HIFs from inductive load increases, such as a motor start or the connection/reclosing of a transformer to a distribution network.
The inductive load increase in complex coordinates is \( Z_{\text{load}} = jX_{\text{load}} \), where \( X_{\text{load}} = \omega_g L_{\text{load}} \), and in \( dq \) coordinates is 
\begin{equation*}
Z_{\text{load}}(s) = \begin{bmatrix} s L_{\text{load}} & -\omega_g L_{\text{load}} \\ \omega_g L_{\text{load}} & s L_{\text{load}} \end{bmatrix}.
\end{equation*}

The poles for each of the four transfer functions after a load increase, obtained with the same parameter values, are
\begin{equation*}\label{loadpol1}
s_{1,2,\text{load}} = \frac{-3}{200L_{\text{load}+30}} \; \pm \; j.
\end{equation*}

As $s_{1,2,\text{fault}} \neq s_{1,2,\text{load}}$, we expect that, given sufficient data, an rARX model should be able to distinguish between an HIF and an inductive load.

\end{document}